\documentclass{article}
\usepackage{amsmath}
\usepackage{amsfonts}
\usepackage{amssymb}
\usepackage{graphicx}
\usepackage{float}
\usepackage{subfigure}
\usepackage{amsmath}
\usepackage{braket}
\usepackage[T1]{fontenc}
\usepackage{textcomp}
\usepackage{gensymb}
\def\ra{\rangle}

\def\bcen{\begin{center}}
\def\ecen{\end{center}}
\begin{document}
\bibliographystyle{unsrt}
\bcen
{\Large Quantum interference induced photon localisation and delocalisation in coupled cavities}\\
\vspace{0.5in}
Nilakantha Meher and S. Sivakumar\\Materials Physics Division\\ 
Indira Gandhi Centre for Atomic Research\\ Kalpakkam 603 102 INDIA\\
Email: siva@igcar.gov.in
\ecen
\begin{abstract}

 We study photon localisation and delocalisation in a system of two nonlinear cavities with intensity-dependent coupling.  It is shown that  complete localisation or delocalisation is possible for proper choices of the strengths of nonlinearity, detuning and inter-cavity coupling.   Role of the relative phase in the initial superposition in attaining localisation and  delocalisation is discussed. Effects of dissipation and decoherence  are considered and the use of quantum interference in reducing dissipation is explored.  Many of the features of the system are shown to be the consequences of quantum interference.  
\end{abstract}

\section{Introduction}
Interference is a consequence of superposition.   Occurrence of interference pattern in Young's double slit experiment, lasing without inversion \cite{Har1}, 
electromagnetically induced transparency \cite{Har2,Bol}, coherent population trapping \cite{Alz}, Hong-Ou-Mandel interferometer\cite{Hong}, quantum phase transition\cite{Green}  {\it etc},  are all consequences of interference  of transition amplitudes making the phenomenon of interference ubiquitous. There are other situations of interest  such as photon localisation in cavities \cite{Sch}, ideal single photon sources \cite{Tang},  {\it etc.},  where quantum interference plays a major role. Ideal single photon sources \cite{Brah}-\cite{Yuan} have  been achieved in cavity QED {\it via} quantum anharmonicity ladder in the energy spectrum due to the presence of strong non-linear interaction\cite{Birn}. In the context of two coupled cavities, localisation corresponds to having  all the photons in one of the cavities, while delocalisation  corresponds to nearly equal number of photons in both the cavities.  These systems are of  special significance as they are realizable and many experiments have been reported in the literature \cite{Chun}-\cite{Sarc}.   \\  

    Photon transport in coupled cavity arrays \cite{Ogd} is dependent on the inter-cavity coupling and the nonlinearity present in the cavities \cite{Albe}.  Cavity nonlinearlity is achieved using suitable material medium in the cavity.   Availability of a material medium within the cavity brings about interesting phenomena.  For instance, in the presence of a two-level atom in the middle cavity of a system of three cavities, complete single photon transfer is forbidden if the atom-field coupling is rotating-wave approximated \cite{Fel}.  Localisation and delocalisation of photons can occur in circuit QED which generates ultrastrong matter-field coupling between superconducting qubits and microwave photons in a resonator \cite{Sch}.  In all these studies, local nonlinearity and inhomogeneous inter-cavity couplings have been employed in achieving localisation and  delocalisation \cite{Sara}.  Another phenomenon that occurs due to strong nonlinear interactions is photon blockade\cite{Ima} which has been observed in cavity QED \cite{Birn, Ada}, ultra-strong coupling regime in atom-field interaction \cite{Rid} and  optomechanical systems\cite{Rabl}. \\

In the present work,  a system of two coupled cavities is considered. Analyses known in the literature are mostly in the context of linear Jaynes-Cummings interaction between cavities \cite{Ogd, Fel} which contain  non-linear Kerr medium \cite{Albe}.  As a generalisation,  inter-cavity coupling is considered to be intensity-dependent,
 which is the quantum equivalent of nonlinearly coupled classical oscillators.   Also, the cavities are considered to have  nonlinear medium, in particular, Kerr medium.  A special form of deformed algebra appears as a natural choice  in studying the dynamics of the system.  We study the localisation and delocalisation phenomena of two photons in the system. If the two photons are detected in one of the cavities after a measurement, it corresponds to  two photon localisation (TPL). On measurement, if each cavity is found to have  one photon each, it is  two-photon delocalisation(TPD). Emergence of these features are understood in terms of quantum interference.\\

   The organization of the article is  as follows.  The generalised Hamiltonian and the dynamics generated by it are discussed in section II.  Importance of the relative phase in the initial state, inter-cavity detuning and nonlinearity  for  TPL and TPD is studied in Section III.  Effects of dissipation and dephasing on TPL and TPD are explored Section IV.  In Section V, linearly coupled system of N cavities is considered and the probability of TPD and TPL is discussed. Also, emergence of two photon trapping in a system of $N$ linearly coupled cavities is discussed to stress the roll of relative. Main results are summarised in Section VI.

\section{Hamiltonian for nonlinearly coupled cavities}
In this section, a system of two ideal cavities is described. The Hamiltonian for the system, setting $\hbar=1$, is  
\begin{multline}\label{Hamilt}
H=\omega_1 a_1^{\dagger}a_1+\omega_2 a_2^{\dagger}a_2+\chi _1 a_1^{\dagger 2} a_1^2+\chi_2 a_2^{\dagger 2}a_2^2+\\
J\left[\sqrt{1+ka_1^{\dagger}a_1}a_1a_2^{\dagger}\sqrt{1+ka_2^{\dagger}a_2}+a_1^{\dagger}\sqrt{1+ka_1^{\dagger}a_1}\sqrt{1+ka_2^{\dagger}a_2}a_2\right]. 
\end{multline}
Here $a_m$ and $a_m^\dagger$ are the annihilation and creation operators for the two cavities $(m=1,2)$.   The first two terms correspond to  independent linear cavities.  The next two terms which depend on $\chi_1$ and $\chi_2$ account for the Kerr nonlinearity in the cavities.  The last term containing the coupling constant $J$ describes an intensity-dependent interaction between the two cavities.  Such interaction terms have been considered in the context of intensity-dependent atom-field coupling \cite{Siva}.\\

The purpose of studying the system described by $H$ is that many other well known interactions are special cases of $H$. In the limit of vanishing $k$, $H$ describes the well known Jaynes-Cummings type model for two cavities,
\begin{align*}
H_{JC}=\omega_1 a_1^{\dagger}a_1+\omega_2 a_2^{\dagger}a_2+J(a_1a_2^{\dagger}+a_1^{\dagger}a_2).
\end{align*}  \\
For instance,  models such as Buck-Sukumar \cite{Buck} and Kerr Hamiltonian\cite{Agar, Yurk, Kir} are obtained if $k\langle a^\dagger a\rangle>>1$ and $k\langle a^\dagger a\rangle<<1$ respectively.\\

To discuss the general case, consider the following deformed operators: $K_m=\sqrt{1+ka_m^{\dagger}a_m}a_m$, and  $K_m^{\dagger}=a_m^{\dagger}\sqrt{1+ka_m^{\dagger}a_m}$ where $m=1$ and 2 correspond respectively to the first and second cavities. The commutator $[K_m,K_m^\dagger]=2K_0$, with $K_0=ka^{\dagger}a+\frac{1}{2}$,  which is identity when $k=0$. These deformed operators form a closed algebra, with Heisenberg-Weyl  and SU(1,1) as limiting cases in $k$ \cite{Siva}.  Under the action of these operators, the number states transform as follows,
\begin{align*}
K\ket{n}&=\sqrt{n}\sqrt{1+k(n-1)}\ket{n-1},\\
K^\dagger\ket{n}&=\sqrt{1+kn}\sqrt{n+1}\ket{n+1}.
\end{align*}
 With $\chi_m=\omega_m k$, the Hamiltonian $H$ is re-expressed in terms of the deformed operators to yield 
\begin{align}\label{HamiltK}
H=\omega_1 K_1^{\dagger}K_1+\omega_2 K_2^{\dagger}K_2+J(K_1K_2^{\dagger}+K_1^{\dagger}K_2).
\end{align} 
This has the same form as the usual JC model. 
Though the Hamiltonian includes intensity-dependent interaction as well as Kerr non linearity,  it is still possible to identify a constant of motion, namely, 
the operator corresponding to the number of quanta $N=a_1^{\dagger}a_1+a_2^{\dagger}a_2$ so that  $[H, N]=0$. 
Existence of this constant of motion implies that there are invariant subspaces for the unitary dynamics generated by $H$.\\

To discuss the occurrence of TPL and TPD, the two cavities are initially prepared in  superposition,
\begin{equation*}
\ket{\psi (0)}=C_1(0)\ket{20}+C_2(0)\ket{11}+C_3(0)\ket{02}.
\end{equation*}
Each of the superposed state in the initial state has two quanta. Therefore, the initial state belongs to the invariant subset spanned by  eigenstates of $N$ with eigenvalue 2.  As a consequence, the time-evolved state also belongs to the invariant subset.  On unitary evolution under $H$, the  state $\ket{\psi}$ evolves into
\begin{equation}
\ket{\psi (t)}=C_1(t)\ket{20}+C_2(t)\ket{11}+C_3(t)\ket{02},
\end{equation}
where 
\begin{align*}
C_1(t)&=((a^2+b^2)L_1-L_2a+L_3)C_1(0)+(L_1(ab+bc)-L_2b)C_2(0)+L_1b^2C_3(0),\\
C_2(t)&=(L_1(ab+bc)-L_2b)C_1(0)+(L_1(2b^2+c^2)-L_2c+L_3)C_2(0)\\
&+(L_1(bc+bd)-L_2b)C_3(0),\\
%\hbox{and}~
C_3(t)&=L_1b^2C_1(0)+(L_1(bc+bd)-L_2b )C_2(0)+(L_1(b^2+ d^2)-L_2d+L_3)C_3(0).
\end{align*}
Various terms occurring in the coefficients are 
\begin{align*}
L_1&=\frac{e^{\lambda_1 t}}{(\lambda_1-\lambda_2)(\lambda_1-\lambda_3)}+\frac{e^{\lambda_2 t}}{(\lambda_2-\lambda_1)(\lambda_2-\lambda_3)}+\frac{e^{\lambda_3 t}}{(\lambda_3-\lambda_1)(\lambda_3-\lambda_2)}, \\
L_2&=\frac{e^{\lambda_1 t}(\lambda_2+\lambda_3)}{(\lambda_1-\lambda_2)(\lambda_1-\lambda_3)}+\frac{e^{\lambda_2 t}(\lambda_3+\lambda_1)}{(\lambda_2-\lambda_1)(\lambda_2-\lambda_3)}+\frac{e^{\lambda_3 t}(\lambda_2+\lambda_1)}{(\lambda_3-\lambda_1)(\lambda_3-\lambda_2)}, \\
L_3&=\frac{e^{\lambda_1 t}(\lambda_2\lambda_3)}{(\lambda_1-\lambda_2)(\lambda_1-\lambda_3)}+\frac{e^{\lambda_2 t}(\lambda_3\lambda_1)}{(\lambda_2-\lambda_1)(\lambda_2-\lambda_3)}+\frac{e^{\lambda_3 t}(\lambda_2\lambda_1)}{(\lambda_3-\lambda_1)(\lambda_3-\lambda_2)},
\end{align*}
where $\lambda _1,\lambda _2$ and $\lambda _3$ are the eigenvalues of $-iH$ in the two photon subspace. Here, \\
$a=-i2\omega_1(1+k)$, $b=-i\sqrt{2(1+k)}J $, $c=-i(\omega_1+\omega_2)$ and  $d=-i2\omega_2(1+k)$. \\\\
Choosing $C_1(0)=\cos\theta$, $C_2(0)=0$ and $C_3(0)=e^{i\phi}\sin{\theta}$ for the initial state,

\begin{equation}\label{initial state}
\ket{\psi}=\ket{\theta,\phi}=\cos{\theta}\ket{20}+e^{i\phi}\sin{\theta}\ket{02},
\end{equation} 
makes it a TPL state, that is, on detection both the photons will be  in only one of the two cavities.  
For this choice of the initial state, the coefficients in the time-evolved state are
\begin{align*}
C_1(t)&=((a^2+b^2)L_1-L_2a+L_3)\cos\theta+L_1b^2e^{i\phi}\sin\theta,\\
C_2(t)&=(L_1(ab+bc)-L_2b)\cos\theta+(L_1(bc+bd)-L_2b)e^{i\phi}\sin\theta,\\
C_3(t)&=L_1b^2\cos\theta+(L_1(b^2+ d^2)-L_2d+L_3)e^{i\phi}\sin\theta.
\end{align*}
%We make special choices of  $\theta$ and the relative phase $\phi$ so as %to achieve maximum  TPL or TPD. 
For use in the subsequent discussions, we define
\begin{eqnarray}\label{Entstate}
\ket{+}&=\ket{\theta=\frac{\pi}{4},\phi=0}=\frac{1}{\sqrt{2}}(\ket{20}+\ket{02})\nonumber\\
\ket{-}&=\ket{\theta=\frac{\pi}{4},\phi=\pi}=\frac{1}{\sqrt{2}}(\ket{20}-\ket{02})
\end{eqnarray}
It may be noted that the state $\vert +\ra$ is symmetric under the exchange of photons while $\vert -\ra$ is antisymmetric.  
   
\section{Localisation and delocalisation}

	To discuss in quantitative terms about TPL and TPD, relevant probabilities are defined:   
the probability of detecting the system to be in $\vert 02\ra$ or $\vert 20\ra$ is  the localisation probability and  
the probability of detecting the system in the state $\vert 11\ra$  is the delocalisation probability. Perfect localisation corresponds to the probability being unity for detecting two photons in one of the cavities and zero for the other. Interpreted in terms of the average number of photons, $\langle a^{\dagger 2} a^2\rangle=2$ for the cavity with two photons and zero for the other. Considering these, it is prudent to choose the  
zero time delay second-order correlation function ($g^{(2)}(0)$),  defined as 
\begin{equation*}
g^{(2)}_i(0)=\frac{\langle a_i^\dagger a_i^\dagger a_i a_i\rangle}{\langle a_i^\dagger a_i\rangle^2},~~~~~~~(i=1,2)
\end{equation*}
as a quantitative tool for distinguishing TPD and TPL .\\

	If $P_{\ket{20}}(P_{\ket{02}})$ is the probability of detecting two (zero) photons in the first cavity and zero (two) photons in the second cavity, then $P_{\ket{02}}+P_{\ket{20}}$ is the TPL probability. Similarly, $P_{\ket{11}}$ is the TPD probability which corresponds to  detecting one photon each in both the cavities corresponding to TPD.\\ 

 The time-evolved state given in the previous section is used to calculate the required probability amplitudes. For the state $\ket{\psi}$ given in the Eqn. \ref{initial state} with $\theta=\frac{\pi}{4}$, the probability of TPD is
\begin{align}\label{Deloc}
P_{\ket{11}}=\vert C_2\vert^2=\frac{|b|^2}{2} \left | \left[(1+e^{i\phi})\left[L_2+iL_1(\omega_1+\omega_2)\right]+2iL_1(1+k)(\omega_1+e^{i\phi}\omega_2)\right] \right |^2
\end{align}
\\  
	
	To understand the origin of two-photon localisation and delocalisation, consider the following transition amplitudes:\\
probability amplitude for transition from $\ket{20}$ to $\ket{11}$ is $C_{\ket{20}\rightarrow\ket{11}}=L_1(ab+bc)-L_2b$,\\
probability amplitude for transition from $\ket{02}$ to $\ket{11}$ is $C_{\ket{02}\rightarrow\ket{11}}=L_1(bd+bc)-L_2b$.\\
These two transition amplitudes are of equal magnitude at resonance but differ in phase by $\phi$, which is the relative phase in the initial state defined in Eqn. \ref{initial state}.   Therefore \cite{Fey}, 
\begin{equation}\label{DestInt}
P_{\ket{11}}(t)=\vert C_{\ket{20}\rightarrow\ket{11}}+e^{i\phi}C_{\ket{02}\rightarrow\ket{11}}\vert^2.
\end{equation}

\begin{figure}[H]
\centering
\includegraphics[scale=0.3]{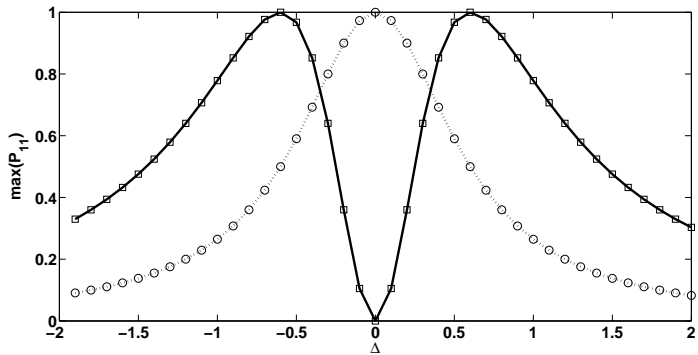}
\caption{Maximum value of probability of delocalisation $P_{\ket{11}}$ as a function of cavity detuning $\Delta$.  The two curves correspond to  two different initial states, namely, $\ket{+}$ (dotted line) and $\ket{-}$ (continuous).   Parameters chosen are $J=0.3$ and $k=0$.  }
\label{fig:Fig1}
\end{figure}

	If the relative phase $\phi=0$ implying  $\ket{\psi}$=$\ket{+}$ and the cavities are resonant $(\Delta=\omega_1-\omega_2=0)$, then $P_{\ket{11}}$ varies between 0 and 1 periodically. On the other hand,  if $\phi=\pi$, i.e.,  $\ket{\psi}$=$\ket{-}$, then  
\begin{align*}
P_{\ket{11}}(t)&=\vert L_1b \vert^2\vert a-d \vert^2\\
&=\vert L_1b \vert^2\vert (-2i\omega_1+2i\omega_2) \vert^2(1+k)^2.
\end{align*}
If the two frequencies are equal ($\omega_1=\omega_2$), then $P_{\ket{11}}$ is $0$ during time-evolution and its holds  for all values of  $J$ and $k$.   In Fig. \ref{fig:Fig1}, the maximum 
achievable $P_{\ket{11}}$ is shown as a function of the detuning parameter when  nonlinearity is absent $(k=0)$. \\

\begin{figure}[H]
\centering
\includegraphics[scale=0.3]{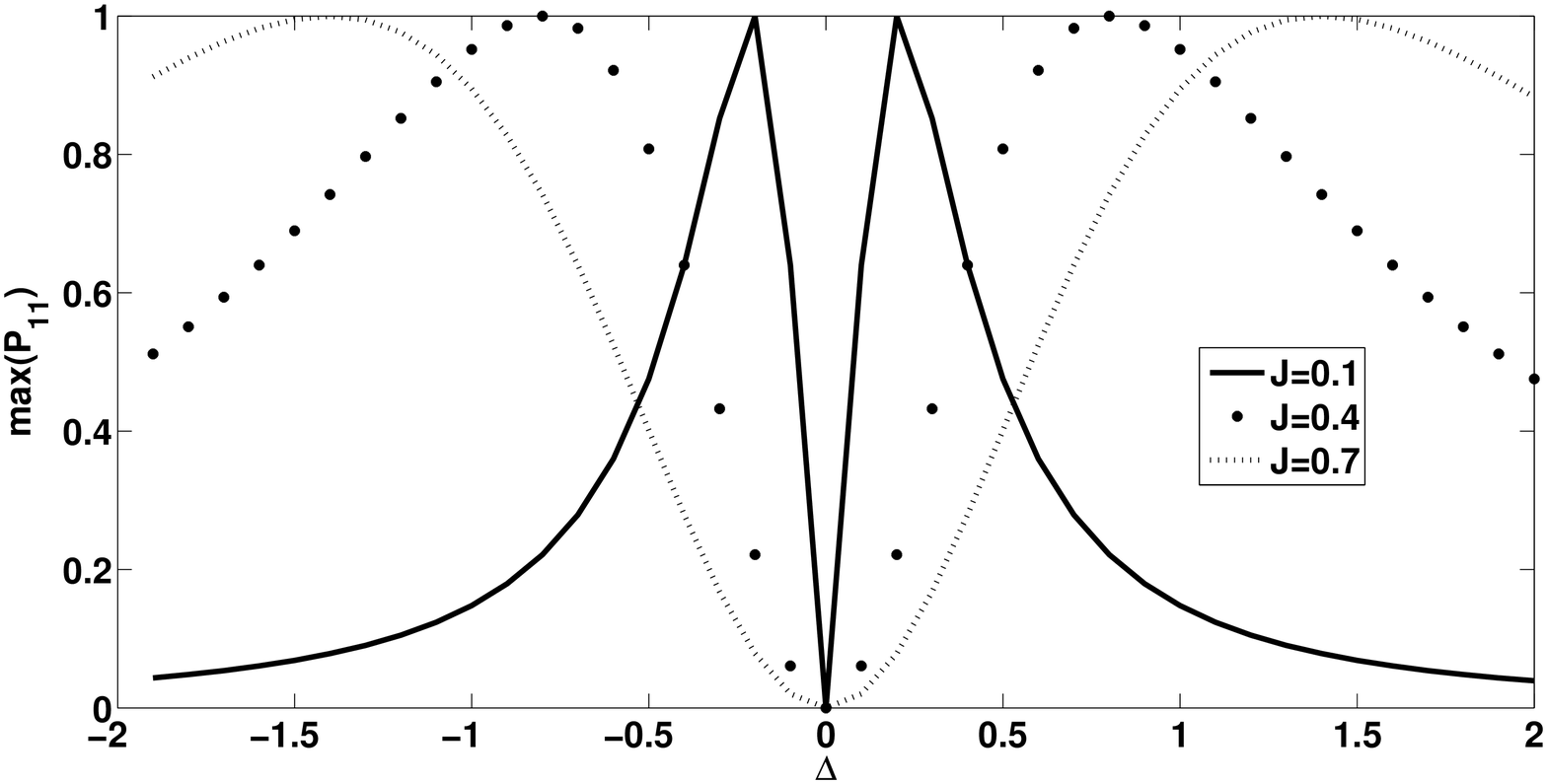}
\caption{Maximum of the probability of delocalisation $P_{\ket{11}}$ with cavity detuning $\Delta$ for the state $\ket{-}$ when nonlinearity is absent ($k=0$). Complete delocalisation occurs at $\Delta=\pm 0.2$, $\pm 0.8$, $\pm 1.4$ for $J=0.1$, $0.4$ and $0.7$ respectively.}
\label{fig:Fig2}
\end{figure}

 In essence, for the state $\ket{+}$ $(\phi=0)$, constructive interference between the two transition amplitudes   enhances the probability of detecting $\ket{11}$ state which is TPD state.  For the state $\ket{-}$ $(\phi=\pi)$, due to destructive interference between the two amplitudes, the probability of detecting the system in the state $\ket{11}$ vanishes.  It is to be pointed out that the initial state  $\ket{-}$ is an eigenstate of $H$ under resonance. As a consequence, the state does not change during evolution apart from overall phase factor. \\
The discussion so far has been restricted to the resonant case. If $\Delta\ne 0$,  then the maximum attainable delocalisation probability $P_{\ket{11}}$ for the state $\ket{+}$ decreases with increasing  detuning as seen from Fig. \ref{fig:Fig1}.  However, the state  $\ket{-}$ evolves to attain complete delocalisation due to constructive interference if $|\Delta|=2J$. For $k=0$ and $\phi=\pi$, Eqn. \ref{Deloc} becomes
\begin{equation*}
P_{\ket{11}}=(2J\Delta)^2|L_1|^2
\end{equation*} 
    Note that $P_{\ket{11}}=0$ if for $\Delta=0$ (destructive Interference) or $J=0$ (the cavities are not coupled). Occurrence of complete delocalisation for a particular value of detuning prompts the question of deciding the right combination of $J$ and $\Delta$ to attain $P_{\ket{11}}\approx 1$.  In Fig. \ref{fig:Fig2}, the maximum value of the delocalisation probability for the state $\ket{-}$ is shown as a function of detuning for different choices of the coupling strength $J$.  It is found that the relation $\vert\Delta\vert=2J$ holds for other values of $J$ too.   If  $(\vert\Delta\vert\ne 2J)$,  delocalisation probability is less than unity. 
%%%%%%%%%%%%%%%%%%%%%%%%%%%%%%%%%%%%%%%%
%%%%%%%%%%%%%%%%%%%%%%%%%%%%%%%%%
  Considering the initial state to be one of the product states  $\ket{\psi}=\ket{20}$ or $\ket{\psi}=\ket{02}$ obtained by setting $\theta=0$ or $\pi/2$ in Eq. \ref{initial state} respectively,  the corresponding delocalisation probabilities are,
\begin{align}
P_{\ket{11}}(\theta=0)&=|b|^2|(iL_1(\omega_1(3+2k)+\omega_2)+L_2)|^2\\
P_{\ket{11}}(\theta=\frac{\pi}{2})&=|b|^2|(iL_1(\omega_2(3+2k)+\omega_1)+L_2)|^2.
\end{align}
In the linear case $(k=0)$,  max($P_{\ket{11}}$) does not exceed $\frac{1}{2}$ at resonance.  On detuning, probability of localisation of two photons is more than the delocalisation probability if  $|\Delta|>2J$.\\

\begin{figure}[H]
\centering
\includegraphics[scale=0.3]{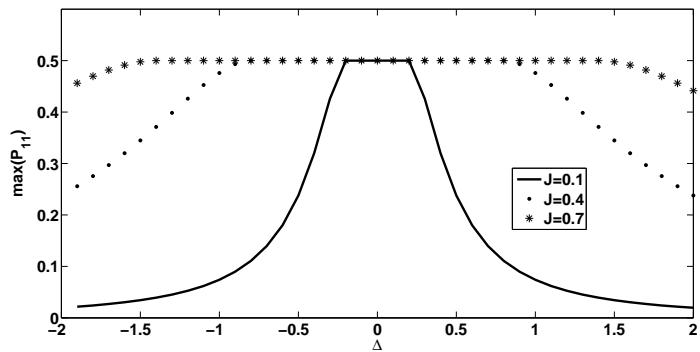}
\caption{Maximum value of $P_{\ket{11}}$ with cavity detuning $\Delta$ for $\ket{20} $ with $k=0$. TPD probability starts decreasing at $\Delta=\pm 0.2$, $\pm 0.8$, $\pm 1.4$ for $J=0.1$, $0.4$ and $0.7$  respectively. It is clear that for $|\Delta|>2J$, TPL dominates over TPD. }
 \label{fig:Fig3}
\end{figure}

 In Fig. \ref{fig:Fig3},  variation of maximum 
delocalisation probability is shown as a function of the detuning parameter $\Delta$ in the linear case. It is seen that max($P_{\ket{11}}$) remains constant for $\vert\Delta\vert\leq2J$ in each case: $J=0.1$ (solid), $0.4$ (dot), $0.7$ (star).
%%%%%%%%%%%%%%%%%%%%%%%%%%%%%%%%%%%%%%%%%%%%%%%%%%%%%%%%%%%%%%%%%%%%%%%%%%%%%%%%%%%%%%%%%%%%%%%%%%%%%%%

In short, for the initial state which is a  localised product state,  TPL dominates over TPD if $|\Delta|>2J$.
%%%%%%%%%%%%%%%%%%%%%%%%%%%%%%%%%%%%%%%%%%%%%%%%%%%%%%%%%%%%%%%%%%%%%%%%%%%%%%%%%%%%%%%%%%%%%%%%%%%%%%%%
\subsection{Role of Nonlinearity}

	The discussions in the last subsection indicate that complete localisation and delocalisation occur if the initial state is a two photon entangled state with appropriate phase evolving under linear JC-type interaction of two coupled cavities.  The role of quantum interference between the relevant transition amplitudes has been emphasized.   Presently, the role of nonlinearity in achieving TPL and TPD is discussed.\\  

\begin{figure}[H]
\centering
\includegraphics[scale=0.3]{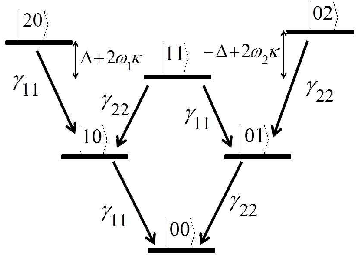}%JPG
\caption{Energy levels  are labelled by the expectation value of the Hamiltonian $H$ in the respective states. The inter-state decay rates are denoted by $\gamma$ with appropriate suffixes.}
\label{fig:FigEnergy}
\end{figure}

Considering  the Hamiltonian in Eqn. \ref{HamiltK}, its expectation values in the relevant two photon states are given by 
\begin{eqnarray}\label{AvgEng}
\bra{20}H\ket{20}&=&2\omega_1(1+k),\nonumber\\
\bra{11}H\ket{11}&=&(\omega_1+\omega_2),\nonumber\\
\bra{02}H\ket{02}&=&2\omega_2(1+k).
\end{eqnarray}
If $\Delta=0$ (resonance) and $(k=0)$ (linear JC), these expectation values are equal. In the presence of nonlinearity $(k\neq0)$, the average energies of $\ket{20}$ and $\ket{02}$ shift by $\Delta+2k\omega_1$ and $-\Delta+2k\omega_2$ respectively from the expectation value in  the state $\ket{11}$ as shown in Fig. \ref{fig:FigEnergy}.  In this figure, the energy levels have been arranged by  their average energies \cite{Adam}.

For the localised states $\vert 20\ra$ or $\vert 02\ra$ to evolve to become the delocalised state $\vert 11\ra$, the respective average energies should be the same in the localised and delocalised states, that is, the states are degenerate in terms  of their expectation values.    This modification of expectation values  is achievable by detuning the cavities depending on the initial state as specified here.  The required detuning has to be $\Delta=-2\omega_1k$ if the initial state is $\ket{\psi}=\ket{20}$ and  $\Delta=2k\omega_2$ for $\ket{\psi}=\ket{02}$.  These relations holds good if $k>J$ \cite{Sara}. When the system is detuned for one of the transitions, the other transition does not occur.  For example, if detuning is appropriate for $\vert 20\ra\rightarrow$ $\ket{11}$ $\left [\bra{20}H\ket{20}=\bra{11}H\ket{11} \right ]$, transition to $\ket{02}$ does not occur as the average energy in the state has shifted by $4\omega_1 k(k+1)$.  Thus, in the presence of Kerr nonlinearity, detuning can be used as a switch to block $\vert 20\ra\rightarrow\vert 02\ra$ transition.  Interestingly, at resonance, transitions from $\vert 11\ra$ to $\vert 02\ra$ or $\vert 20\ra$ and vice-versa are nearly forbidden if $k >> J$(Figure. \ref{fig:FigEnergy}).  What happens in this limit is that the presence of a photon in a cavity blocks the inflow of photon from the other cavity, which is analogous to the photon blockade phenomenon in a driving single cavity. This stabilizes the $\vert 11\ra$ state. \\ 

If $\vert +\ra$ or $\vert -\ra$ is the initial state, then, 
in presence of nonlinearity, the state  does not evolve to  have complete overlap with $\vert 11\ra$ for any detuning.   In fact, under resonance the initial state $\ket{-}$  remains as a  TPL state  as inferred from  Eq. \ref{DestInt} even though nonlinearity could be present.\\ 

If the intensity-dependent coupling is absent, delocalisation probability can attain its  maximum in the presence of Kerr nonlinearity under appropriate  detuning.  If the initial state is $\ket{\psi}=\ket{20}$, the required detuning is  $\Delta=-2\chi_1$ while $\Delta=+2\chi_2$ if $\ket{\psi}=\ket{02}$ is the initial state. This conditions arises on demanding  equal average energies of the states in the Kerr Hamiltonian \cite{Agar, Yurk, Kir}.

\section{TPD in presence of dissipation and dephasing}

An ideal cavity is characterized by complete isolation from the influences of  the environment.
 In reality, however, there are unavoidable influences from the environment leading to dissipation and decoherence or dephasing.   The dominant mechanism of dissipation is photon leakage. Dephasing is another aspect of system-environment interaction which leads to decay of the off-diagonal elements of the density operator. In this process the system loses its quantum coherence but not energy.
  
\subsection{TPD in presence of dissipation}          

Effects of dissipation is studied by analyzing the  master equation for the density operator of the system.  It is to be pointed out that in the presence of dissipation the previously considered invariant subset of the Hilbert space is inadequate as the number of photons is not fixed.  However, since  the dissipative process does not increase the number of quanta, only states with  lower number of quanta than what is contained in the initial state are to be considered.  To facilitate writing down the master equation, relevant states are relabeled as follows: 
$\ket{00}\rightarrow\ket{1}\rangle,\ket{10}\rightarrow\ket{2}\rangle,\ket{01}\rightarrow\ket{3}\rangle,\ket{20}\rightarrow\ket{4}\rangle,\ket{11}\rightarrow\ket{5}\rangle,\ket{02}\rightarrow\ket{6}\rangle$, where double angular brackets are used to represent the various bipartite states of the two cavities. Using these as the basis vectors,  the elements of the density operator are obtained by solving  the master equation \cite{Agarwal,Gard, Carm, Fic},
\begin{align}\label{MasEqn}
\frac{\partial\rho}{\partial t}&=-i[H,\rho]+\sum\limits_{i,j=1}^2\frac{\gamma_{ij}}{2}(2a_j \rho a_i^\dagger-a_i^\dagger a_j \rho -\rho a_i^\dagger a_j).
\end{align}
 Here $\gamma_{11}$ and $\gamma_{22}$ are decay rates of the first and second cavities respectively 
and, $\gamma_{12}$ and $\gamma_{21}$ are the cross-damping rates arising  due to interference of transition amplitudes. \\

If $\gamma_{12}=\gamma_{21}=0$  and $\gamma_{11}=\gamma_{22}=\gamma$, 
the localisation probability is
\begin{equation*}
P_{\ket{20}+\ket{02}}(t)=e^{-2\gamma t}(|C_1(t)|^2+|C_3(t)|^2).
\end{equation*}
The result shows that the TPL probability  falls exponentially in time at a rate that is twice the decay rate of the cavities.   The suffix $\ket{20}+\ket{02}$ indicates probability corresponds to the case when the state of the system subsequent to measurement involves localised two photon states.\\

\begin{figure}[H]
\centering
\includegraphics[scale=0.3]{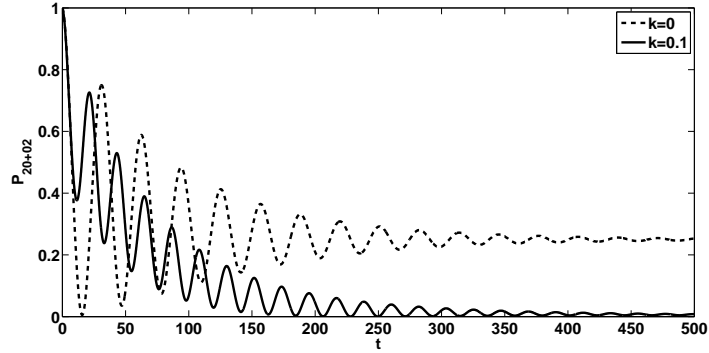}
\caption{Probability of localisation $P_{\ket{20}+\ket{02}}=\rho_{44}+\rho_{66}$ is shown as a function of $t$ for the initial $\ket{+}$.  Values uses are $J=0.05$,$\gamma=0.005$ and $\Delta=0$. Her $\gamma=\gamma_{11}=\gamma_{22}=\gamma_{12}=\gamma_{21}$.
Continuous curve corresponds to the linear case $k=0$ and the dashed line for intensity-dependent interaction with $k=0.1$.  }
\label{fig:Fig7}
\end{figure}

To bring out the effects of cross-damping, the master equation is solved numerically to get $\rho _{44}$ and $\rho _{66}$. Referring to the convention given in the beginning of the subsection, it is immediate that $\rho _{44}$ and  $\rho _{66}$ are  the respective probabilities for detecting the system in $\vert 20\ra$ and $\vert 02\ra$.  Therefore, $\rho _{44}+\rho _{66}$  is the localisation probability when system evolves to a mixed state due to dissipation. Evolution equations for these two elements of the density matrix  are 
\begin{align*}
\dot{\rho} _{44}&=-i[J\sqrt{2(1+k)}(\rho _{54}-\rho _{45})]-2\gamma_{11}\rho _{44}-\frac{\gamma_{12}}{\sqrt{2}}\rho_{54}-\frac{\gamma_{21}}{\sqrt{2}}\rho_{45},\\
\dot{\rho} _{66}&=-i[J\sqrt{2(1+k)}(\rho _{56}-\rho _{65})]-2\gamma_{22} \rho _{66}-\frac{\gamma_{12}}{\sqrt{2}}\rho_{65}-\frac{\gamma_{21}}{\sqrt{2}}\rho_{56},
\end{align*}
where $\rho_{ij}=\langle\bra{i}\rho\ket{j}\rangle$.\\
Temporal evolution of TPL probability in presence of dissipation is shown in Fig. \ref{fig:Fig7}.  The initial state is $\ket{+}$.   Both linear JC model($k=0$) and intensity-dependent interaction ($k=0.1$) have been considered.  Due to interference between the various  transitions  shown in Fig. \ref{fig:FigEnergy}, the probability of localisation does not completely vanish  in the absence of nonlinearity. Including nonlinearity $(k\ne 0)$  leads to loss of interference leading to complete decay of localisation.   Essentially, nonlinearity destroys the interference between the amplitudes by shifting the average energies in the states  $\ket{20}$ and $\ket{02}$ from $\ket{11}$. Also, detuning makes the average energy of all the components unequal, which leads to asymmetry in the system and destroys the coherence between the various transition.  As a result, localisation probability decays to zero. 

In classical dissipative systems, energy  decays to zero if there is no external pumping.  So, it is important  to study the corresponding situation in the present system.  Using the master equation given in Eq. \ref{MasEqn}, the expectation values of products of creation and annihilation operators of the two linear cavities are obtained.  These equations can be cast in the following form, 
\begin{equation}
\frac{d}{dt}\left(\begin{array}{c} \langle a_1^\dagger a_1\rangle\\\langle a_1^\dagger a_2\rangle\\\langle a_1 a_2^\dagger\rangle\\\langle a_2^\dagger a_2\rangle\end{array}\right)=\left(\begin{array}{cccc}
-\gamma   & \frac{-iJ}{\hbar}-\frac{\gamma}{2}&\frac{iJ}{\hbar}-\frac{\gamma}{2}&0  \\
\frac{-iJ}{\hbar}-\frac{\gamma}{2}    & -\gamma &0 & \frac{iJ}{\hbar}-\frac{\gamma}{2}\\
\frac{iJ}{\hbar}-\frac{\gamma}{2} &0 & -\gamma & \frac{-iJ}{\hbar}-\frac{\gamma}{2}\\
0&\frac{iJ}{\hbar}-\frac{\gamma}{2}&\frac{-iJ}{\hbar}-\frac{\gamma}{2} & -\gamma
\end{array}\right)\left(\begin{array}{c} \langle a_1^\dagger a_1\rangle\\\langle a_1^\dagger a_2\rangle\\\langle a_1 a_2^\dagger\rangle\\\langle a_2^\dagger a_2\rangle\end{array}\right).\\\\ 
\end{equation}
Here 
$\gamma=\gamma_{11}=\gamma_{22}$ and $\gamma_{12}=\gamma_{21}=\sqrt{\gamma_{11}\gamma_{22}}$ \cite{Bos}.\\ 

The matrix differential equation is solved to get the average number of photons in the first cavity.  The resultant expression is 
\begin{align*}
\langle a_1^\dagger a_1 \rangle _t&=\frac{1}{4}[X_1(t)\langle a_1^\dagger a_1 \rangle _0+X_2(t)\langle a_1^\dagger a_2 \rangle _0+X_3(t)\langle a_1 a_2^\dagger \rangle _0+X_4(t)\langle a_2^\dagger a_2 \rangle _0],
\end{align*}
where
\begin{align*}
X_1(t)&=e^{\lambda_1 t}+e^{\lambda_2 t}+e^{\lambda_3 t}+e^{\lambda_4 t},\\
X_2(t)&=-e^{\lambda_1 t}+e^{\lambda_2 t}-e^{\lambda_3 t}+e^{\lambda_4 t},\\
X_3(t)&=e^{\lambda_1 t}-e^{\lambda_2 t}-e^{\lambda_3 t}+e^{\lambda_4 t},\\
X_4(t)&=-e^{\lambda_1 t}-e^{\lambda_2 t}+e^{\lambda_3 t}+e^{\lambda_4 t},
\end{align*}
and 
\begin{equation}
\lambda_1=2iJ-\gamma,~\lambda_2=-2iJ-\gamma,~
\lambda_3=0,\lambda_4=-2\gamma.
\end{equation}
Similar expression can be derived for the average number of photons in the second cavity.\\

For the initial state $\ket{\psi}=\cos\theta\ket{20}+e^{i\phi}\sin\theta\ket{02}$ ,  the average photon number in the cavities is
\begin{align*}
\langle a_1^\dagger a_1 \rangle _t=\frac{1}{2}[(e^{\lambda_1 t}+e^{\lambda_2 t})cos2\theta
+e^{\lambda_3 t}+e^{\lambda_4 t}].
\end{align*} 
It is clear from the expression that the evolution of the mean photon number is independent of the relative phase.\\ 

	For $\theta=\pi/4$,
\begin{equation}
\langle a_1^\dagger a_1 \rangle _t=\langle a_2^\dagger a_2 \rangle _t=\frac{1}{2}[e^{-2\gamma t}+1],
\end{equation}
which saturates at $1/2$ for large t.  Thus, quantum interference  stabilizes the average number photons at a non-zero value in spite of dissipation.
If the initial state is $\ket{\psi_+}=\frac{1}{\sqrt{2}}\left[\ket{10}+\ket{01}\right]$, then   
\begin{equation*}
\langle a_1^\dagger a_1 \rangle _t=\langle a_2^\dagger a_2 \rangle_t=\frac{e^{-2\gamma t}}{2}.
\end{equation*}
The average number of photons in the cavities will decay to zero as $t$ increases.  Both the cavities lose energy at the same rate  a consequence of assuming resonance and equal damping.  If the initial state is $\ket{\psi_-}=\frac{1}{\sqrt{2}}(\ket{10}-\ket{01})$, the average number of photons is  
\begin{equation*}
\langle a_1^\dagger a_1 \rangle _t=\langle a_2^\dagger a_2 \rangle_t=\frac{1}{2}.
\end{equation*} 
The average number of photons saturates at 1/2.  That the average does not decay to zero is due to the destructive interference between the amplitudes corresponding to $\ket{10}\rightarrow\ket{00}$ and $\ket{01}\rightarrow\ket{00}$ as a consequence of the relative phase $\pi$ in the initial state. 
The rate of loss of photons depends on the initial state, especially through its dependence on $\theta$, refer Eq. \ref{initial state}.  In the present context,  the initial state  is either the symmetric state $\vert +\ra$  or the antisymmetric state $\vert -\ra$. The former decays at a the rate $2\gamma$ while the later does not decay.   This is analogous to the superradiance and subradiance  that occur in the interaction between a three-level atom  and field \cite{Fic}. Due to interference, the average photon number saturates at 1/2  though dissipation is present.  It implies that quantum interference makes it possible to retain nonzero number of photons in the cavities in spite of dissipation. 
%%%%%%%%%%%%%%%%%%%%%%%%%%%%%%%%%%%%%%%%%%%%%
\subsection{TPD in presence of dephasing}
 As seen in the previous section, relative phase in the initial state plays a crucial role in achieving TPD.  In the presence of dephasing, the relative phases in the evolved state are randomized in time.  It is natural to expect dephasing to affect TPD whose occurrence is sensitive to the relative phase.    
The master equation described in the previous section can be modified to incorporate dephasing by  including the Lindblad term 
$\frac{\gamma_d}{2}\mathcal{D}(a^\dagger a)\rho$ \cite{Tang,Car}.  With this modification, the master equation becomes\\
\begin{equation*}
\frac{\partial\rho}{\partial t}=-i[H,\rho]+\frac{\gamma_d}{2}\mathcal{D}[a_1^\dagger a_1]\rho+\frac{\gamma_d}{2}\mathcal{D}[a_2^\dagger a_2]\rho,
\end{equation*}
where the operator $D[a^\dagger a]\rho$ is 
\begin{equation*}
\mathcal{D}[\hat{o}]\rho=2\hat{o}\rho\hat{o}^\dagger-\hat{o}^\dagger\hat{o}\rho-\rho\hat{o}^\dagger\hat{o}.
\end{equation*}
This Lindblad term accounts for dephasing  which leads to the decay of  the off-diagonal elements in the density operator.\\

\begin{figure}[H]
\centering
\includegraphics[scale=0.3]{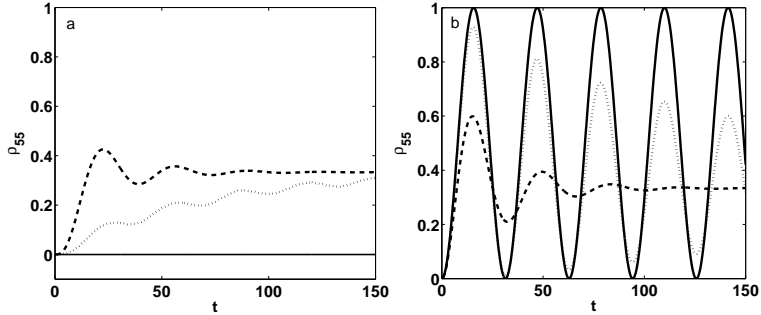}
\caption{Probability of delocalisation $\rho_{55}$ in the linear case $(k=0)$ as a function of $t$ for $(a)$ $\ket{-}$ and $(b)$ $\ket{+}$. Curves corresponds to different values of decay rate: $\gamma_d=0$(continous),  $0.005$(dot) and $0.05$(dash).  For all the cases,  $J=0.05$ and $\Delta=0$.}
\label{fig:Fig8}
\end{figure}

	To understand the effect of dephasing, the numerically obtained time-dependence of  $\rho_{55}$, which is the probability of delocalisation $(P_{\ket{11}})$, is  shown in Fig. \ref{fig:Fig8} . If the initial state is $\ket{-}$, the time evolved state is partially delocalised as $\rho_{55}$ is less than unity. In the absence of dephasing, complete localisation is possible due to destructive interference between the amplitudes corresponding to the transitions  $\ket{20} \rightarrow \ket{11}$ and $ \ket{02} \rightarrow \ket{11}$.  But dephasing randomizes the  relative  phases during evolution and suppresses the destructive interference.
Generally, in the presence of dephasing, the initial coherence is expected to vanish resulting in an equilibrium density operator Fig. \ref{fig:Fig8}.  The steady-state  density matrix elements  are 
\begin{equation*}
\rho_{44}=\frac{1}{3},
\rho_{45}=0,
\rho_{46}=0,
\rho_{55}=\frac{1}{3},
\rho_{56}=0,
\rho_{64}=0,
\rho_{66}=\frac{1}{3}.
\end{equation*} 
It is also clear from Fig. \ref{fig:Fig8} that the shown curves indeed saturate at 1/3.\\    

\begin{figure}[H]
\centering
\includegraphics[scale=0.3]{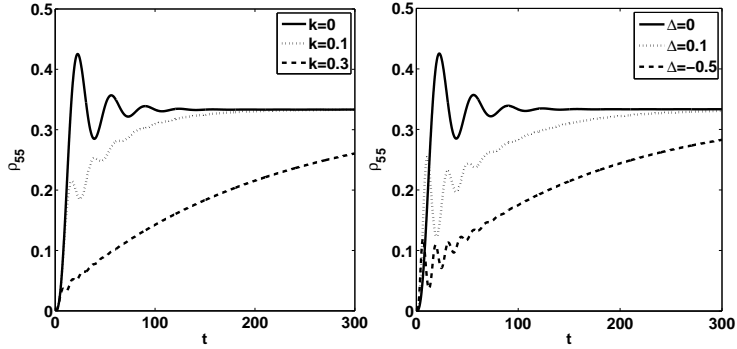}
\caption{Delocalisation probability as a function of $t$ for  $\ket{-}$. (a) With various $k$. Continuous line is for $k=0$, dotted line is 
for $k=0.1$ and dashed line is for $k=0.3$ with $J=0.05$, $\Delta=0$ and $\gamma_d=0.05$.(b) With various detuning. Continuous line for $\Delta=0$, dotted line 
is for $\Delta=0.3$ and dashed line correspond to $\Delta=-0.5$ with $J=0.05$, $k=0$ and $\gamma_d=0.05$. Net effect of nonlinearity, detuning and dephasing on $\ket{+}$ is same i.e non-linearity and detuning slow down the process of attaining equilibrium density operator.  }
\label{fig:Fig9}
\end{figure}

The probability of TPD for different values of the nonlinearity parameter  $k$ are shown in Fig. \ref{fig:Fig9}(a).  In Fig. \ref{fig:Fig9}(b), the  probability for TPD  is shown for various values of  $\Delta$.  Comparing the curves corresponding to different values of $\Delta$ it is clear the rate  of attaining the equilibrium values is slow.  Similar, conclusions are drawn by comparing the curves corresponding to different values of $k$ in Fig. \ref{fig:Fig9}(a).  Both detuning and nonlinearity slow down the process of attaining steady state.
%%%%%%%%%%%%%%%%%%%%%%%%%%%%%
\subsection{Role of coherence in delocalisation}

\begin{figure}[H]
\centering
\includegraphics[scale=0.3]{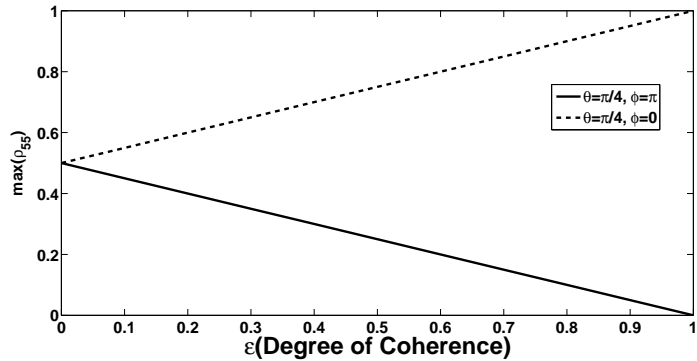}
\caption{Maximum probability of delocalisation $\rho_{55}$ as a function of $\epsilon$. Continuous line corresponds to state $\ket{-}$ 
and dotted line corresponds to $\ket{+}$.  Other parameters are  $J=0.05$, $k=0$, $\Delta=0$. 
}
 \label{fig:Fig10}
\end{figure}

   For a  better appreciation of the role of coherence, consider the realistic situation where an initial pure state is prepared with probability $\epsilon$ 
and a related random state (noise) with probability $1-\epsilon$.  With initial TPL state $\ket{\psi}$ and the added noise M, the total density matrix is \cite{R}, 

\begin{equation}\label{DegCoh}
\rho=\epsilon\ket{\psi}\bra{\psi }+(1-\epsilon)M,
\end{equation}
where 
\begin{align*}
\vert\psi\ra&=\cos\theta\ket{20}+e^{i\phi}\sin{\theta}\ket{02}\\
M&=\cos^2\theta\ket{20}\bra{20}+\sin^2\theta\ket{02}\bra{02},
\end{align*}
This is a mixed state for all $\epsilon<1$.
The state interpolates between the TPL state $\ket{\psi}$ which has coherence  and the state $M$ which has no 
coherence.  Thus, $\epsilon$ measures the degree of coherence in the state $\rho$.  
To bring out the effect of initial coherence in the state given in Eqn. \ref{DegCoh}, the variation of maximum of TPD probability with $\epsilon$ is shown in Fig. \ref{fig:Fig10}. The value of  $\theta$ is $\frac{\pi}{4}$, which corresponds to equal magnitude of the superposition coefficients in the initial state.
The curves shown in the figure correspond  to two values of $\phi$, namely, $0$ and $\pi$.   In the later case, as $\epsilon$ increases the peak value of TPD probability $\rho_{55}$ decreases and vanishes at $\epsilon=1$. This is due to the destructive interference between the amplitudes for the two transitions, namely $\ket{20}\rightarrow\ket{11}$ and $\ket{02}\rightarrow\ket{11}$. In the former case, the peak of TPD probability increases with $\epsilon$ due to constructive interference. The initial state with full coherence ($\epsilon=1$) evolves to attain perfect TPD and TPL.

\section{Role of entanglement}
The discussion in the preceding sections have been focused on the role of the  initial phase and the coherence on the emergence of TPD and TPL. Another important aspect to be considered is the entanglement in the initial state.  In fact, the state $\ket{+}$ and $\ket{-}$ are  entangled states. If concurrence is used as the measure of entanglement for bipartite qutrit, it is clear that both $\ket{+}$ and $\ket{-}$ have equal entanglement. The quantity of concurrence for the state $\ket{\psi}$ in Eqn. \ref{initial state} is \cite{Cer}
\begin{equation*}
C(\theta)=\sqrt{3|\sin\theta\cos\theta|^2},
\end{equation*}
For both the states $\ket{+}$ and $\ket{-}$, $C=\sqrt{3}/2$. Though the two states have equal entanglement, the state $\ket{+}$ can evolve to attain complete TPD while $\ket{-}$ does not when the cavities are resonant. The main lesson of this example is that the relative phase plays a stronger  role than entanglement.   This observation is at variance with the claim that entanglement is the only parameter required to quantify TPD and with increase of the former the later increases too\cite{Shi}.   To discuss further, consider the case of N-coupled cavity array. \\ 

	The Hamiltonian for  N identical cavities that are linearly coupled is $(\hbar=1)$,
\begin{equation*}
H=\omega \sum\limits_{j=1}^N a_j^\dagger a_j +J \sum\limits_{j=1}^{N-1}(a_j^{\dagger} a_{j+1}+H.c).
\end{equation*}
This Hamiltonian can be written in uncoupled form using normal mode operators leading to the following diagonal form:
\begin{equation*}
H=\sum\limits_{k=1}^N \Omega_k  c_k^\dagger c_k, ~~~~~~\Omega_k =\omega+2J\cos(\frac{\pi k}{N+1}).
\end{equation*}
Using Heisenberg equation for the time-development of the operators, in particular, the annihilation operator, leads to \\
\begin{equation*}
a_j(t)=\sum_{i} G_{jl}(t) a_l(0),
\end{equation*}
\begin{equation*}
G_{jl}(t)=\sum_{k=1}^N e^{-i\Omega_k t} S(j,k) S(l,k),
\end{equation*}
\begin{equation*}
S(j,k)=\sqrt{\frac{2}{N+1}}\sin(\frac{j\pi k}{N+1}).
\end{equation*}
Two photon states are considered once again, nevertheless, with the additional freedom that the two photons can be shared by any pair of cavities, say, $r$ and $s$, 
among the $N$ cavities.  The corresponding localised state is $\ket{\psi}=\cos\theta\ket{2}_r\ket{0}_s+e^{i\phi}\sin\theta\ket{0}_r\ket{2}_s$.   
 Specifically, concurrence vanishes  for $\theta=0,\pi/2$ and $C=\sqrt{3}/2$ for $\theta=\pi/4$(independent of $\phi$).\\
  
	To assess the role of the relative phase in inducing TPD or TPL in the array, the  
joint probability $P_{mn}$ of coincidence detection of two photons in the  cavities $m$ and $n$ is considered.  In the present case, it is 
defined as  
\begin{equation*}
P_{m,n}=\langle a_n^{\dagger}(t)a_m^{\dagger}(t)a_m(t)a_n(t)\rangle.
\end{equation*}
%The probability to detect both the photons in the same cavity is %$P_{n,n}(t)/2$.\\ 

\begin{figure}[H]
\centering
\includegraphics[scale=0.43]{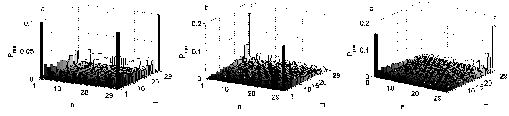}%JPG
\caption{The joint probability for the two photon coincidence detection in $m$th and $n$th cavity at time t=83.57 with J=0.1.(a)for $\theta=0$, $C=0$, (b) 
for $\theta=\pi/4$, $\phi=0$, $C=\sqrt{3}/{2}$,(c) for $\theta=\pi/4$, $\phi=\pi$, $C=\sqrt{3}/{2}$.}
\label{fig:Fig11}
\end{figure}

	For the  initial state $\ket{\psi}$, 
\begin{align}
P_{mn}=&2|\cos\theta G_{mr}(t)G_{nr}(t)+e^{i\phi}\sin\theta G_{ms}(t)G_{ns}(t)|^2.
\end{align}
Assuming $r=15$ and $s=16$ and $N=29$, correlation function calculated from the previous expression is shown in Fig. \ref{fig:Fig11}.
%%%%%%%%%%%%%%%%%%%%%%%%%%%%%%%%%%%%%%%%%%%%%%%%%%%%%%%%%%%%%%%%%%%%%%%%%%%%%%%%%%%%%%%%%%%%%%%%%%%%%%%%%%%%%%%%%%%%%%%%%%%%%%%%
%%%%%%%%%%%%%%%%%%%%%%%%%%%%%%%%%%%%%%%%%%%%%%%%%%%%%%%%%%%%%%%%%%%%%%%%%%%%%%%%%%%%%%%%%%%%%%%%%%%%%%%%%%%%%%%%%%%%%%%%%%%%%%%%%

\begin{figure}[H]
\centering
\includegraphics[scale=0.3]{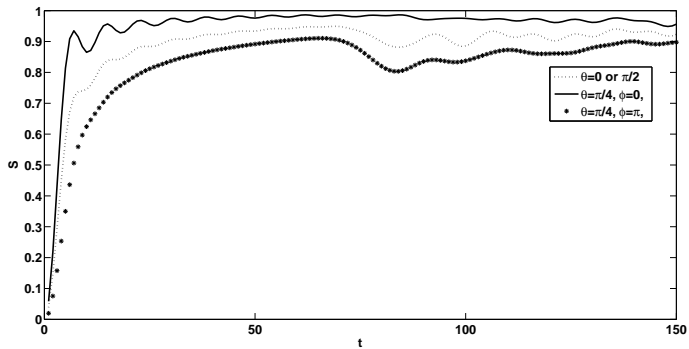}
\caption{Time evolution of degree of two photon delocalisation $S$ for different initial states with J=0.1.}
\label{fig:Fig12}
\end{figure}

The quantities  $P_{mn}$ can be treated as the elements of a correlation matrix.
Each diagonal element represents the probability of coincidence detection of two photon in same cavity which is localisation.  Each off-diagonal element represents the probability of  delocalisation. Comparing Fig. \ref{fig:Fig11}(b) and \ref{fig:Fig11}(c) corresponding to two different states with same initial concurrence, it is very clear that 
the former has maximum delocalisation and the later has comparatively less 
delocalisation.   A simpler way of seeing this is to calculate  the degree of TPD ,\\
\begin{equation*}
S=1-\frac{1}{2}\sum_{n=1}^N P_{n,n}(t)
\end{equation*}
which defines the probability of detection of two photon in two different cavity. If S=0, then it is TPL state and if S=1, then it is TPD state. Evolution of degree of delocalisation for different initial entangled states are shown in Fig. \ref{fig:Fig12} and the following are observed.
%%%%%%%%%%%%%%%%%%%%%%%%%%%%%%%%%%%%%%%%%%%%%%%%%%%%%%%%%%%%%%%%%%%%%%%%%%%%%%%%%%%%%%%%%%%%%%%%%%%%%%%%%%%%%%%%%%%%%%%%%%%%%%%%%%
If $\theta=\pi/4$ and $\phi=\pi$, the  concurrence in the corresponding state is $\sqrt{3}/{2}$.  This state  does not evolve to a completely  delocalised state where as the state with C=0 evolves to become a  TPD state.  
The initial phase difference between the two states $\ket{20}$ and $\ket{02}$ in the superposition  is to be considered as an additional parameter  apart from the entanglement to qualify whether a state will lead to delocalisation or not. 
\\
In N linearly coupled cavity arrays, it is interesting to find a localised state, which will never become delocalised  due to destructive interference between various paths of transition i.e two photons completely trapped in any one of the cavity. Such a state is,
\begin{equation*}
\ket{\psi}=\frac{1}{\sqrt{N}}\sum_{n=1}^N(-1)^{(n+1)}\ket{n}
\end{equation*} 
where $n$ indicates the two photons are in the $n$th cavity and other cavities are in vacuum state.

\section{Summary}

   A generalization of the Jaynes-Cummings model to include intensity-dependent coupling of two nonlinear cavities has been introduced. The dynamics of this coupled cavity system allows for  the possibility of two-photon localisation and delocalisation.  Occurrence of these phenomena depends on the relative phase and entanglement in the initial superposed state. The importance of the relative phase in the initial superposition is brought out in the explanation of the localisation and delocalisation as consequences of quantum interference between the transition amplitudes.  Delocalisation occurs when the interference is constructive.   Destructive interference leads to localisation.  This argument  is applicable to any pair of cavities even if there are $N$ coupled cavities. If the initial state   of the two cavities is a localised product state such as $\vert 02\ra$ or $\vert 2 0\ra$, localisation dominates over delocalisation if the detuning is more than a critical value, namely, twice the coupling strength between the cavities. \\  

   Due to the nonlinearity of the cavity, the energy levels of the cavities shift.  The average energies of the localised and delocalised states are different. As a consequence, transition from localised state to delocalised state is blocked at resonance.  In order for the transition to occur from localised state to delocalised state, the average energies of them are to be nearly equal and this can only happen for the localised product states by adjusting strength of nonlinearity and detuning.\\  

   Decoherence due to interaction with the environment prevents complete localisation or delocalisation in the system.  This is consistent with the fact that decoherence affects the relative phase which is very crucial for  the interference of probability amplitudes  to  occur.  In fact, the state $\ket{02}+\ket{20}$ which has the same coherence as the state $\ket{02}-\ket{20}$, evolves to  a TPD state.  But the state $\ket{02}-\ket{20}$ does not evolve and remains a localized state.  The difference in their evolutions is due to the relative phase in the initial state. \\

   In the presence of dissipation,  it always possible for the photons to leak from the system. This allows the system to evolve into a mixed state involving lower number states.  As a consequence, the probability of localisation is less. It may be noted that due to quantum interference,  the system does not reach the vacuum state if the initial state is suitably chosen. \\

              Incorporating nonlinearity, intensity-dependent coupling provides additional control parameters apart from detuning in realizing delocalisation or localisation.   This will be of value in generating the localised or delocalised states for different applications and observing the quantum interference phenomenon in cavity arrays.


\begin{thebibliography}{40}
\bibitem{Har1}{S. E. Harris, Phys. Rev. Lett. {\bf {62}}, 1033 (1989).}
\bibitem{Har2}{S. E. Harris, J. E. Field, and A. Imamo\u{g}lu, Phys. Rev. Lett.
{\bf 64}, 1107 (1990).}
\bibitem{Bol}{K. J. Boller, A. Imamo\u{g}lu, and S. E. Harris, Phys. Rev.
Lett. {\bf 66}, 2593 (1991).}
\bibitem{Alz}{G. Alzetta, A. Gozzini, L. Moi, and G. Orriols, Nuovo Cimento
B {\bf 36}, 5 (1976).}
\bibitem{Hong}{C. K. Hong, Z. Y. Ou, and L. Mandel, Phys. Rev. Lett.{\bf 59}, 2044 (1987).}
\bibitem{Green}{A. D. Greentree, C. Tahan,J. H. Cole, and L. C. L. Hollenberg, Nature Phys.{\bf 2}, 856 (2006).}
\bibitem{Sch}{S. Schmidt \textit{et.al.}, Phys. Rev. B {\bf 82}, 100507 (2010).}
\bibitem{Tang}{Tang \textit{et.al.}, Scientific Reports {\bf 5}, 9252 (2015). }
\bibitem{Brah}{Brahim Lounis and Michel Orrit, Rep. Prog. Phys. {\bf 68}, 1129 (2005).}
\bibitem{McK}{J. McKeever \textit{et.al.}, Science {\bf 303}, 1992 (2004).}
\bibitem{Yuan}{Z. Yuan \textit{et.al.}, Science 295, 102 (2002).}
\bibitem{Birn}{K. M. Birnbaum \textit{et.al.}, Nature (London) {\bf 436}, 87 (2005).}
\bibitem{Chun}{Chun-Wang Wu \textit{et.al.} , Physics Letters A  {\bf 376}, 44 (2012).}
\bibitem{Eil}{J. C. Eilbeck \textit{et.al.}, Physica D {\bf 16}, 318 (1985).}
\bibitem{Smer}{A. Smerzi \textit{et.al.}, Phys. Rev. Lett.{\bf 79}, 4950 (1997).}
\bibitem{Alb}{M. Albiez \textit{et.al.}, Phys. Rev. Lett. {\bf 95}, 010402 (2005).}
\bibitem{Levy}{S. Levy \textit{et.al.}, Nature (London) {\bf 449}, 579 (2007).}
\bibitem{Sarc}{D. Sarchi \textit{et.al.}, Phys. Rev. B {\bf 77}, 125324 (2008).}
\bibitem{Ogd}{C. D. Ogden, E. K. Irish, and M. S. Kim, Phys. Rev. A {\bf 78}, 063805 (2008).}
\bibitem{Albe}{Alberto Biella \textit{et.al.} Phys. Rev. A {\bf 91}, 053815 (2015).}
\bibitem{Fel}{S. Felicetti, G. Romero, D. Rossini, R. Fazio, and E. Solano,
Phys. Rev. A {\bf 89}, 013853 (2014).}
\bibitem{Sara}{ Sara Ferretti \textit{et.al.} Phys Rev A {\bf 82}, 013841 (2010).}
\bibitem{Ima}{A. Imamo\u{g}lu, H. Schmidt, G. Woods, and M. Deutsch, Phys.
Rev. Lett. {\bf 79}, 1467 (1997); P. Grangier, D. F. Walls, and K. M.
Gheri, \textit{ibid}. {\bf 81}, 2833 (1998).}
\bibitem{Ada}{Adam Miranowicz \textit{et.al.} Phys. Rev. A 87, 023809 (2013).}
\bibitem{Rid}{A. Ridolfo, M. Leib, S. Savasta, and M. J. Hartmann, Phys.
Rev. Lett. 109, 193602 (2012).}
\bibitem{Rabl}{ P. Rabl, Phys. Rev. Lett. {\bf 107}, 063601 (2011).}
\bibitem{Siva}{ Sivakumar S, Int. J. Theor. Phys. {\bf 43} 2405 (2004)}
\bibitem{Buck}{Buck B. and Sukumar, C. V. Phys. Lett. A {\bf 81}, 132 (1981). }
\bibitem{Agar}{G.S. Agarwal and R.R. Puri, Phys. Rev. A {\bf 39} 2969 (1989). }
\bibitem{Yurk}{B. Yurke and D. Stoler, Phys. Rev. Lett. {\bf 57}, 13 (1986). }
\bibitem{Kir}{G. Kirchmair, B. Vlastakis, Z. Leghtas, S. E. Nigg, H.
Paik, E. Ginossar, M. Mirrahimi, L. Frunzio, S. M. Girvin,
and R. J. Schoelkopf, Nature (London) {\bf 495}, 205 (2013). }
\bibitem{Fey}{Richard Feynman, Robert Leighton, Matthew Sands, \textit{The Feynman Lectures on Physics} {\bf Vol. III} (1963). }
\bibitem{Adam}{Adam Miranowicz, Jiri Bajer, Neill Lambert, Yu-xi Liu, Franco Nori, {\bf arXiv:1506.08622} [quant-ph]}
\bibitem{Agarwal}{G. S. Agarwal, in \textit{Quantum Statistical Theories of Spontaneous Emission and their Relation to other Approaches}, edited
by G. H\"{o}hler, Springer Tracts in Modern Physics, Vol.{\bf 70}
(Springer-Verlag, Berlin, 1974), Sec. 15.B.}
\bibitem{Gard}{C. W. Gardiner, Phys. Rev. Lett. {\bf 70}, 2269 (1993).}
\bibitem{Carm}{H. J. Carmichael, Phys. Rev. Lett. {\bf 70}, 2273 (1993).}
\bibitem{Fic}{ Z. Ficek and S. Swain,\textit{ Quantum Interference and Coherence:
Theory and Experiments} (Springer, New York, 2005).}
\bibitem{Bos}{A. R. Bosco de Magalh\~{a}es and M. C. Nemes,
Phys. Rev. A {\bf 70}, 053825 (2004).}
\bibitem{Car}{F. Caruso, A. W. Chin, A. Datta, S. F. Huelga, and M. B. Plenio,
J. Chem. Phys. {\bf 131}, 105106 (2009).}
\bibitem{R}{ R de J Leon-Montiel \textit{et.al.} Laser Phys. Lett. {\bf 12} 085204 (2015).}
\bibitem{Cer}{J. L. Cereceda, {\bf arXiv: quant-ph/0305043} }
\bibitem{Shi}{ Tang Shi-Qing \textit{et.al.} Chinese Phys. Lett. {\bf 32} 040303 (2015).}
\end{thebibliography}
\end{document}